\documentclass[twocolumn,epjc3]{svjour3}  
\smartqed
\RequirePackage{graphicx}
\RequirePackage[colorlinks,citecolor=blue,urlcolor=blue,linkcolor=blue]{hyperref}
\journalname{Eur. Phys. J. C}
\usepackage[numbers,sort&compress]{natbib}
\usepackage{graphics}
\usepackage{gensymb}
\usepackage{graphicx,color}
\usepackage{bm}
\usepackage{tabularx}
\usepackage{epstopdf}
\usepackage{enumerate}
\usepackage[normalem]{ulem}
\usepackage{verbatim}
\usepackage[table,xcdraw]{xcolor}
\usepackage{amssymb}
\usepackage{setspace}
\begin{document}

\title{Model dependence of the number of participant nucleons and observable consequences in heavy-ion collisions}

\author{Manjunath~Omana~Kuttan\thanksref{e1,a1,a2,a3} \and Jan~Steinheimer\thanksref{e2,a1} \and Kai~Zhou\thanksref{a1} \and Marcus~Bleicher\thanksref{a2,a4} \and Horst~Stoecker\thanksref{a1,a2,a5} 
}

\thankstext{e1}{e-mail: manjunath@fias.uni-frankfurt.de}
\thankstext{e2}{e-mail: steinheimer@fias.uni-frankfurt.de}

\institute{Frankfurt Institute for Advanced Studies, 
D-60438 Frankfurt am Main, Germany \label{a1} \and Institut f\"ur Theoretische Physik, Goethe Universit\"at Frankfurt, Max-von-Laue-Strasse 1, D-60438 Frankfurt am Main, Germany\label{a2} \and Xidian-FIAS international Joint Research Center, Giersch Science Center,
D-60438 Frankfurt am Main, Germany \label{a3} \and Helmholtz Research Academy Hesse for FAIR (HFHF), GSI Helmholtz Center for Heavy Ion Physics, Campus Frankfurt, Max-von-Laue-Str. 12, 60438 Frankfurt, Germany \label{a4} \and GSI Helmholtzzentrum f\"ur Schwerionenforschung GmbH, D-64291
Darmstadt, Germany \label{a5}}

\date{Received: date / Accepted: date}

\maketitle

\begin{abstract}
The centrality determination and the estimated fluctuations of number of participant nucleons $N_{part}$ in Au-Au collisions at 1.23 $A$GeV beam kinetic energy suffers from severe model dependencies. Comparing the Glauber Monte Carlo (MC) and UrQMD transport models, it is shown that $N_{part}$ is a strongly model dependant quantity. In addition, for any given centrality class, Glauber MC and UrQMD predicts drastically different $N_{part}$ distributions. The impact parameter $b$ and the number of charged particles $N_{ch}$ on the other hand are much more correlated and give an almost model independent centrality estimator. It is suggested that the total baryon number balance, from integrated rapidity distributions, can be used instead of $N_{part}$ in experiments. Preliminary HADES data show significant differences to both, UrQMD simulations and STAR data in this respect.

\end{abstract}
\sloppy
\section{Introduction}
The properties of QCD matter at high baryon density and/or temperature are studied at heavy-ion collision (HIC) experiments throughout the world. Several observables such as the collective flow, particle yields and fluctuations have been proposed to study the properties of the hot, dense matter created in these collisions \cite{Stoecker:1986ci,Hofmann:1976dy,Stoecker:2004qu,Stephanov:1998dy,Hatta:2003wn}. These observables strongly depend on the size of the interaction volume which fluctuates event-by-event with the impact parameter ($b$) or the number of participant nucleons ($N_{part}$) of the collision. Both $b$ and $N_{part}$ are not directly measured by the experiments. Therefore, observables such as the charged track multiplicity $N_{ch}$ or hits in a detector $N_{\mathrm{hits}}$ are used as proxy to determine $b$ and $N_{part}$. 

The Glauber Monte Carlo (MC) model \cite{Glauber:1955qq,Glauber:2006gd} is often used to relate the $N_{ch}$ to $b$ and $N_{part}$ of a collision. At very high collision energies (eg. LHC, RHIC beam energies), the Glauber MC provides a reliable approximation of the initial state of heavy-ion collision to characterise the events based on collision centrality. At high beam energies one can safely assume that the only elastically scattered nucleons will basically fly at a very small angle to the beam and not be measured in most detectors' acceptance. In such a scenario comparing the Glauber model with only inelastic cross section with the 'naive' definition of 'at least one scattering' (and in the acceptance) actually makes sense. However, at low and intermediate beam energies ($\sqrt{s_{\mathrm{NN}}} \lessapprox$ 3-10 GeV), several assumptions of the model become questionable. At these energies, the initial energy deposition and entropy production is not nearly instantaneous and the dynamics of the nuclei during the interpenetration and compression stage are relevant in modelling the collision. Early fluid dynamic simulations have even showed a complete disintegration of the spectator fragments due to the created compression wave, even in peripheral collisions \cite{Baumgardt:1975qv} and more recent studies with a hydro-hybrid model \cite{Cimerman:2023hjw}, as well as full transport model simulations show a strong interaction between the stopped and compressed system and the spectators \cite{Reichert:2023eev}, leading to the large directed flow which is observed at low beam energies (see e.g.\cite{Stoecker:1980vf,Brachmann:1999xt,Ohnishi:2017xjg,HADES:2022osk}). This effect is not included in the Glauber MC model. Furthermore, the Glauber MC model also assumes that the elastic interactions among the nucleons can be neglected to map the $N_{part}$ to the measured particle spectra. This is a valid assumption only in mid rapidity, at high beam energies and the elementary, nucleon-nucleon elastic cross section ($\sigma_{el}^{NN}$) becomes negligible compared to the nucleon-nucleon inelastic cross section ($\sigma_{inel}^{NN}$). For collision energies of $\sqrt{s_{\mathrm{NN}}} \lessapprox$ 3 GeV where $\sigma_{el}^{NN}$ and $\sigma_{inel}^{NN}$ are almost equal, such an assumption is not valid anymore. Nucleons undergoing only elastic interactions during the compression and subsequent expansion stages have a significant contribution on the final state spectra. Despite these inconsistencies, the Glauber MC model is used to fit the experimental data using several parameters with little physical interpretation and are then used to estimate centrality, interaction volume and its fluctuations. Therefore, at low collision energies, it is important to systematically test the Glauber MC model as compared to transport models to which experimental data is compared. 

We investigate in detail the differences between the Glauber MC model and the fully dynamical model Ultra Relativistic Quantum Molecular Dynamics (UrQMD) \cite{Bass:1998ca,Bleicher:1999xi,Petersen:2008dd} and the uncertainties that the differences could introduce in experimental analysis. In particular,we analyse Au-Au collisions at 1.23 \textit{A}GeV which were recently measured by the High Acceptance DiElectron Spectrometer (HADES) \cite{HADES:2009aat} collaboration at the SIS-18 accelerator, GSI. HADES uses the Glauber MC model for characterisation of event centrality \cite{HADES:2017def}\footnotemark{}. Besides HADES, the STAR collaboration at BNL's RHIC has recently measured Au-Au collisions at an intermediate energy of $\sqrt{s_{\mathrm{NN}}}=$ 3 GeV. It was shown that the strong model dependence of the volume fluctuations lead to large uncertainties in the analysis of higher order proton cumulants \cite{HADES:2020wpc,STAR:2022qmt}. In this context, the present study will be useful not just for HADES or STAR, but also for all future intermediate energy HIC experiments in understanding the implications of a Glauber MC-based centrality selection for analysing data at intermediate collision energies.

\footnotetext[1]{ To be more precise, the HADES centrality is based on the summed number of hits detected by the TOF and the RPC detectors which is not identical to the number of reconstructed charged tracks at mid-rapidity. However, it was shown \cite{HADES:2017def} that both quantities are strongly correlated and both give the same results for the mean number of participants as well as the impact parameter, in the Glauber MC study, while using the number of hits shows a better resolution. For simplicity we will simply use the number of charged tracks and assume that all our conclusions also apply for the centrality determination with the number of tracks in the TOF and RPC detectors.}

\section{The Glauber MC model}\label{glaubsec}
The Glauber MC model is a simplified description of the initial overlap phase of heavy-ion collisions which treats the collision as a collection of instantaneous, independent inelastic binary scattering of nucleons travelling on straight trajectories. In the Glauber model, $b$ and $N_{part}$ are mapped to the  experimentally measured $N_{ch}$ distributions through a fitting procedure, assuming that the measured multiplicity $N_{ch}$ is proportional to the number of participants $N_{part}$. The Glauber MC fit to the $N_{ch}$ distribution is then used to define centrality classes and the average values of $b$ and $N_{part}$ and their distributions for the events in any given centrality class are used in the analysis of experimental data.

The Glauber MC model is based on the optical interpretation of the initial interpenetration phase in which collision participants are determined solely from the transverse overlap of the nuclei. The energy and baryon number deposition occurs instantaneously along the longitudinal axis. In this work, for implementing the Glauber MC model, the positions of these nucleons are sampled using a nuclear charge density function for which the two parameter fermi (Wood-Saxon) distribution is used
\begin{equation}
\label{2pf}
    \rho(r)=\frac{\rho_{0}}{1+exp(\frac{r-R}{a})}\ ,
\end{equation}

where $\rho_{0}$ is the charge density at the center, R is the radius parameter and a is the diffuseness parameter. Following the values used in \cite{HADES:2017def} for a gold nucleus, we set $\rho_{0}$=1, R=6.554 and a=0.523. The radial component (r) of the positions of the nucleons are then sampled from the probability distribution 
\begin{equation}
P(r)\propto r^{2}\rho(r) \ ,   
\end{equation}
 
and the azimuthal ($\phi$) and polar ($\theta$) components from uniform probability distributions. To avoid overlapping nucleons, a sampled position is resampled if there already exists another nucleon within a distance of less than $d_{min}$= 0.9 fm. This minimal distance mimics the short range repulsion between nucleons and corresponds to roughly twice the hard-core radius of the nucleons. In \cite{behruzthesis} it was shown that varying the $d_{min}=\pm100\%$ from their default value of 0.9 fm results in an error of about 3-4 $\%$ on the total cross section.

Once the projectile and target nuclei are constructed, the impact parameters (b) of the collisions are sampled from the probability distribution 
\begin{equation}
P(b)\propto b~db \ .   
\end{equation}
The sampled impact parameters are in the range from 0 to $b_{max}$, where $b_{max}$ is set to be greater than the sum of the radi of the projectile and the target. The projectile and target nuclei are positioned such that the separation between their center of masses in the transverse plane is the sampled impact parameter.

The Glauber MC treats the binary collision process via a geometric interpretation of the cross section. Two nucleons are marked as participant nucleons, i.e., as to have undergone a collision, if the transverse distance between them ($d_{ij}$), is less than their radi as corresponding to the nucleon-nucleon inelastic cross section
\begin{equation}
\label{cross}
i.e., \pi d_{ij}^{2}\leq \sigma_{inel}^{NN}.
\end{equation}
Based on \cite{HADES:2017def}, the value of $\sigma_{inel}^{NN}$ is set to 23.8 mb. In this way, all nucleons from either projectile or the target, which have undergone at least one inelastic collision  are counted as "participant nucleons", no matter how many collisions they would actually undergo in a transport model (see the discussion on the UrQMD model description in the next section).

The next step in Glauber MC is to map the number $N_{part}$ to experimental observables, usually the number of measured charged particles $N_{ch}$. In the "wounded nucleon" model, $\langle  N_{ch}\rangle $ is assumed to be proportional to $\langle N_{part}\rangle$. With such an assumption, the number of charged particles per participant can be sampled from a Negative Binomial probability Distribution (NBD) whose mean value and width are chosen such that the sampled $N_{ch}$ distribution fits to the experimental data.

The last, important and possibly problematic step, as we will discuss below, is to define centrality percentiles on the sampled $N_{ch}$ distribution and to calculate the mean $b$ and mean $N_{part}$ for each centrality bin from this Glauber model. The values such defined are then used as the expected average impact parameter $\langle b \rangle$ and number of participants $\langle N_{part} \rangle$ together with the fluctuations thereof. Hence, for a given $N_{ch}$, all events in a given centrality bin of the experimental data are grouped together. This is a drastic difference to N-body models like UrQMD as shown in the next section.

In the present study, this is exemplified by fitting the the Glauber model to UrQMD events in order to systematically analyse the large differences between the Glauber MC predictions and the values directly extracted from a detailed theoretical, semi-classical N-body description such as the UrQMD model. 

\section{The UrQMD model}\label{ursec}
UrQMD is a microscopic, non-equilibrium N-body transport model used for event-by-event simulations of heavy-ion collisions \cite{Bass:1998ca,Bleicher:1999xi,Petersen:2008dd}. In this study we used the cascade version of UrQMD 3.5.
In UrQMD, positions of the initial projectile and target nucleons are sampled using a Wood-Saxon distribution similar to the Glauber MC \footnote{Note, that both the Glauber MC and UrQMD model assume the same distribution for protons and neutrons and do not include 2-particle or short-range correlations inside the nucleus.}. The minimal distance in the UrQMD initialization is 1.6 fm which is almost a factor 2 larger than in the Glauber model. But as discussed earlier, this only has a small effect on the total cross section as we will see later. In this case, the sampled nucleons also have a randomly chosen Fermi momentum between 0 and $P^{max}_{F}$ where

\begin{equation}
P^{max}_{F}=\hbar c(3 \pi^{2} \rho)^\frac{1}{3} \ .
\end{equation}

UrQMD, in cascade mode, propagates hadrons and their resonances on straight trajectories. New particles can be produced in the model through resonance decays or string excitation and fragmentation. During the evolution, the transverse separation between every pair of particles($d_{ij}$) is calculated at the point of closest approach and if the geometric area as defined by ($d_{ij}$) is less than the total cross section ($\sigma^{ab}_{tot}$) for the particle species involved, the particles are considered to undergo a collision, i.e.

\begin{equation}
i.e.,\pi d_{ij}^{2}\leq \sigma^{ab}_{tot} \ .
\end{equation}

 Note that in UrQMD, to determine whether a binary collision occurs, the total cross section ($\sigma^{ab}_{tot}=\sigma^{ab}_{inel}+\sigma^{ab}_{el}$) is used, while the Glauber model considers only the inelastic cross section. The $\sigma^{ab}_{tot}$ in UrQMD depends on the center-of-mass energy of the pair as well as on the particles species ($a,b$) involved . The separation of the collision into elastic or inelastic collisions, subsequent particle productions and decays are then performed as per their respective cross sections which are taken either from experimental h-h data \cite{Workman:2022ynf} when available or from effective parameterization based on phase space considerations and continuations of known cross sections, i.e. through the additive quark model. Any nucleon which has undergone at least one elastic or inelastic interaction during the evolution is then considered as a participant nucleon. 
 
\begin{figure}[t]
    \includegraphics[width=0.48\textwidth]{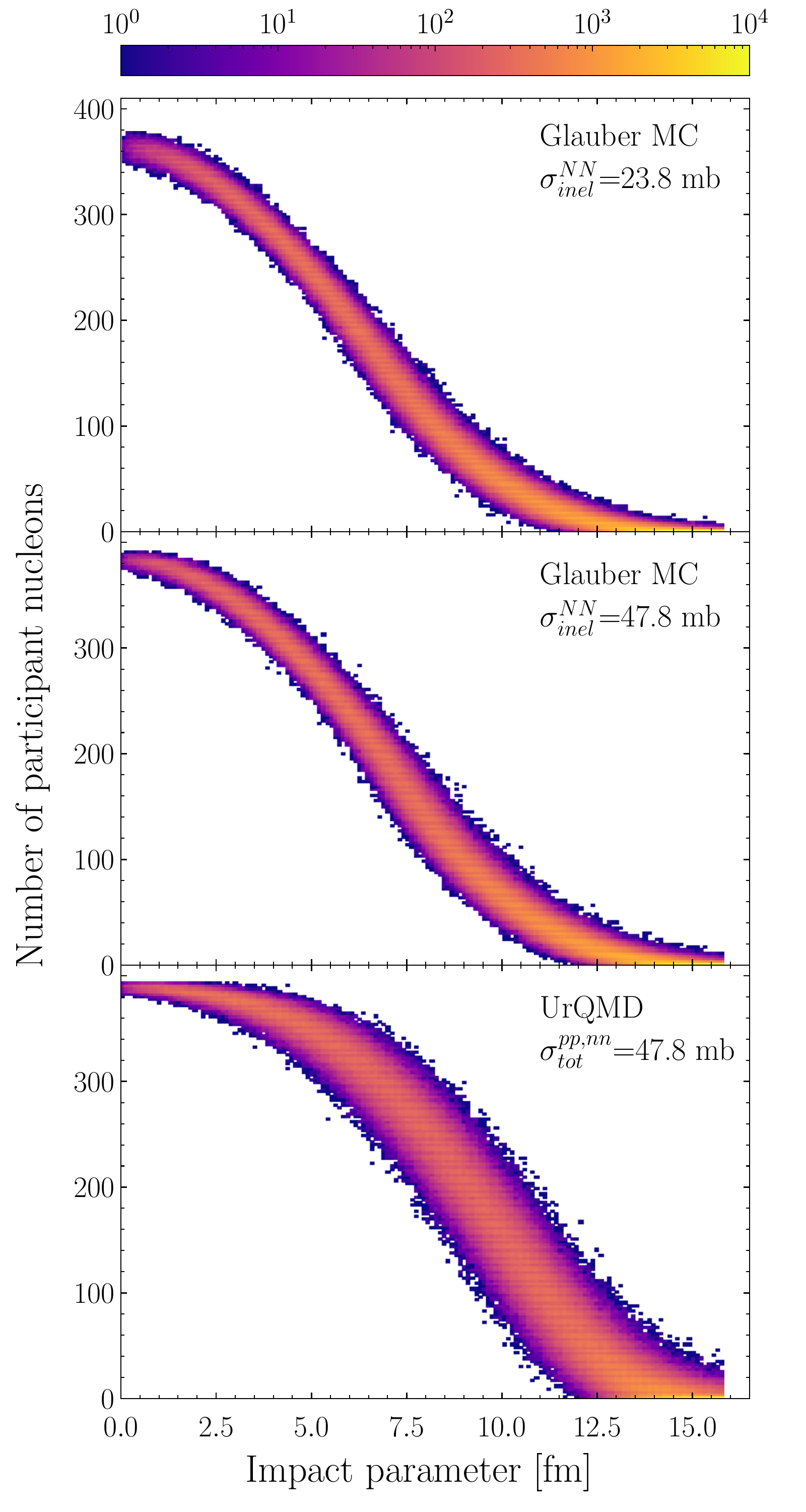}
    \caption{(Color online) $N_{part}$ Vs. $b$ distributions. The upper two panels show the results from  Glauber MC model with $\sigma_{inel}^{NN}$= 23.8 and 47.8 mb, respectively, while the last panel shows the $N_{part}$ distribution from UrQMD. In the second plot, the total proton-proton or neutron-neutron cross section used by UrQMD (47.8 mb) at 1.23 \textit{A}GeV is used as input cross section to Glauber MC. Strong differences are observed between UrQMD and Glauber MC results even if elastic collisions are considered in modelling $N_{part}$. Each plot is generated from $7.5 \times 10^{5}$ events.}   
    \label{4}
\end{figure}
 
\section{$N_{part}$ and $N_{ch}$: Glauber MC Vs UrQMD}
Au-Au collisions at 1.23 \textit{A}GeV were simulated using the Glauber MC and UrQMD models as discussed in sections \ref{glaubsec} and \ref{ursec}. The $N_{part}$ distributions predicted by the two models are shown in figure \ref{4}. The two upper panels in figure \ref{4} show the Glauber MC results while the bottom panel shows the UrQMD results. In UrQMD any nucleon having experienced at least one scattering is considered a participant. It can be seen that at any given impact parameter, the maximum $N_{part}$ given by Glauber MC is smaller that the maximum $N_{part}$ given by UrQMD. For very central collisions ($b< 3 $ fm), the $N_{part}$ distribution from UrQMD shows less variance as compared to Glauber MC. At the same time, for $b> 3 $ fm the UrQMD $N_{part}$ distribution has very large variance as compared to Glauber MC distribution. 

\begin{figure}[t]
    \centering
    \includegraphics[width=0.49\textwidth]{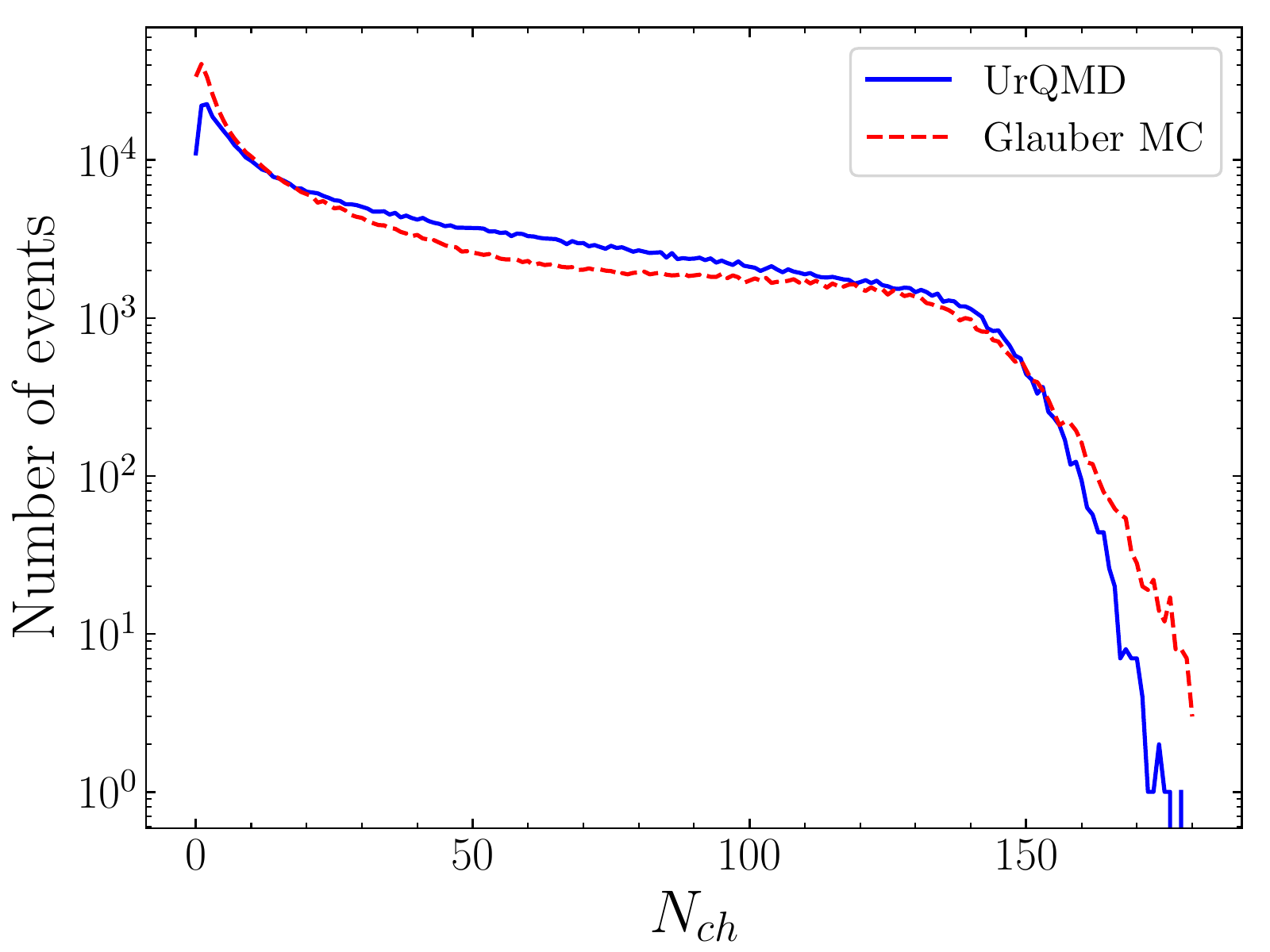}
    \caption{(Color online) Glauber fit to UrQMD cascade data. The UrQMD data comprises of the charged track multiplicity $N_{ch}$ within the HADES acceptance angles (18- 85 \degree) for $6 \times 10^{5}$ inelastic events. The Glauber MC curve is the best fit generated from $6 \times 10^{5}$ inelastic events with $\mu$= 0.4 and k= 30 for eq. \ref{nbd}. The Glauber MC $N_{ch}$ fits reasonably well to UrQMD data for central collisions $N_{ch}>110$  but underestimates the number of events for intermediate collisions ( $25 <N_{ch}< 110$). }   
    \label{0}
\end{figure}

As pointed out earlier, one difference between Glauber MC and UrQMD is that Glauber MC neglects the elastic interactions while UrQMD takes into account both elastic and inelastic nucleon-nucleon interactions. The Glauber MC is a reasonable assumption for very high beam energies, where mostly, newly produced particles contribute  to the $N_{ch}$ distributions. However, at SIS18 energies, this is certainly not the case.
 
To study the effect of including elastic interactions in Glauber MC, we also used the total p-p cross section value taken from UrQMD (47.8 mb) at 1.23 $A$GeV as input to Glauber MC collision criteria (eq. \ref{cross}) instead of just the inelastic cross section. The resulting $N_{part}$ distribution is shown in the middle panel of figure \ref{4}. Including elastic interactions into the Glauber model increases the maximum $N_{part}$ reached in very central collisions and comes close to the UrQMD values. However, the overall width of the distribution has not changed and is still very different and much narrower, for $b>3$ fm, than the UrQMD results. 

It is interesting to see how these differences in $N_{part}$ are manifested when mapped to observables such as the charged track multiplicity. Experimental analyses usually map the Glauber MC generated $N_{part}$ to the measured charged track multiplicity. For a consistent comparison between the models, we map the Glauber MC generated $N_{part}$ to the UrQMD distribution of charged tracks within the HADES acceptance, i.e., between 18 degree to 85 degree polar angle. The $N_{part}$ and $b$ distributions obtained from Glauber MC for the centrality classes as defined by the Glauber MC fit to the $N_{ch}$ distribution are then compared to the true distributions of b and $N_{part}$ from UrQMD simulations. 

To fit the "measured" UrQMD $N_{ch}$ with the Glauber MC model, a given $N_{part}$ is related to the charged track multiplicity per participant ($n$) by a Negative Binomial Distribution (NBD)
 
\begin{equation}
\label{nbd}
P(n)=\frac{\Gamma(n+k)(\mu/k)^{n}}{\Gamma(n+1) \Gamma(k)(\mu/(k+1))^{n+k}} \ .
\end{equation}

The same method is used in the data analysis, where the values for $\mu$ and $k$ are fitted to the measured $N_{ch}$ distribution. They are later used to sample the number of charged tracks produced by each participant nucleon in the event.

The Glauber MC model is fit to the charged track distribution from UrQMD within the HADES acceptance using 600,000 UrQMD events, each with at least one inelastic collision. The result of the fit is shown in figure \ref{0}. The NBD parameters are fit to minimise the deviation from the UrQMD "data" for both central and mid-central collisions (number of tracks $>$90). This corresponds to $\mu$= 0.4 and $k$= 30 in the Glauber MC fit curve (red dashed) as shown in figure \ref{0}. The Glauber MC fit matches the UrQMD data reasonably well for central events, i.e., events with number of tracks $>$110. However the Glauber MC fit overestimates the number of events by an order of magnitude for extremely central events. The Glauber MC fit also underestimates the number of charged tracks for intermediate collisions (25 $<$ number of tracks $<$110) and overestimates it again for extremely peripheral collisions(number of tracks $<$10). In practice, experiments also often fold this multiplicity distribution with further efficiency functions to account for multiplicity dependent uncertainties and other non-linear effects which tend to improve the fit. The objective of this work is to investigate the model dependency of volume fluctuations. However, introducing more parameters with little physical significance to improve the Glauber MC fit would only marginally modify the volume fluctuations in the Glauber model. Therefore, such corrections are not implemented in the present work and the fit results simply indicate that the simplified Glauber MC approach cannot provide an accurate description of collisions as a dynamical model like UrQMD does.

For the ease of reading, in the following we will be referring to the Glauber fit to the UrQMD $N_{ch}$ distribution within HADES acceptance (figure \ref{0}, red curve) as "Glauber-$N_{ch}$" and the true UrQMD $N_{ch}$ distribution within the HADES acceptance (figure \ref{0}, blue curve) as "UrQMD-$N_{ch}$" distribution.

\begin{figure}[t]
    \centering
    \includegraphics[width=0.5\textwidth]{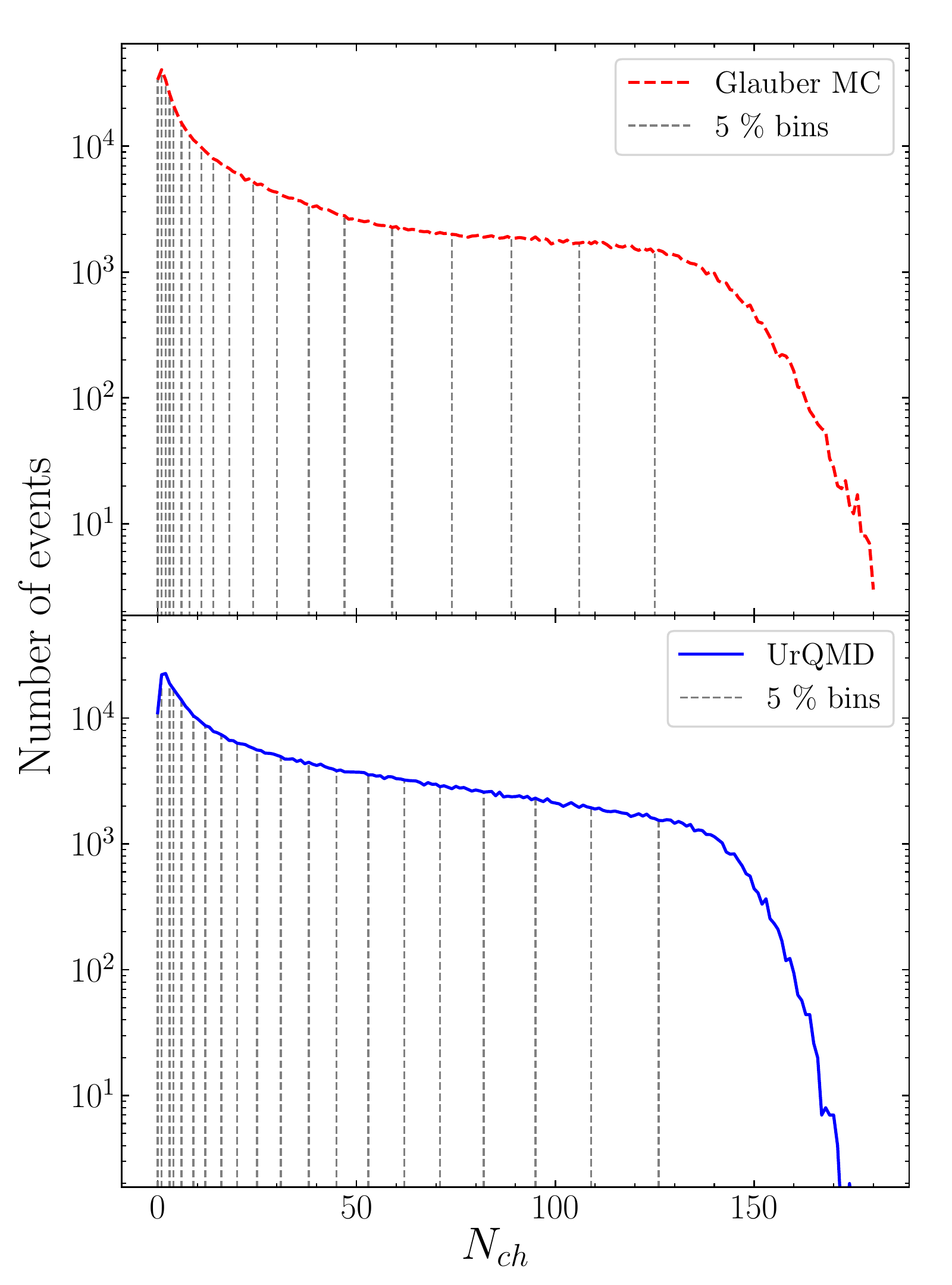}
    \caption{(Color online) 5$\%$ Centrality bins. The top plot shows centrality classes as defined based on Glauber MC fit to UrQMD $N_{ch}$ distribution  while the bottom plot shows centrality classes defined directly on UrQMD $N_{ch}$ distribution. Only $N_{ch}$ within the HADES detector acceptance for inelastic events from UrQMD simulations are considered in both cases. The bin boundaries are marked as dashed vertical lines.} 
    \label{cenbins}
\end{figure}

\section{Centrality selection}
The centrality C for events with $N_{ch}= n$, is defined as the fraction of the total cross section ($\sigma_{tot}$) given by 
\begin{equation}
    C=\frac{1}{\sigma_{tot}}\int_{n}^{N_{ch}^{max}} \frac{d\sigma}{dN_{ch}}   dN_{ch} \ .
    \label{cen}
\end{equation}
 It follows that sharp cuts in $N_{ch}$ based on their percentile score can be used to define different centrality classes. In figure \ref{cenbins}, centrality bins of 5$\%$, defined as sharp cuts in two different $N_{ch}$ distributions, are visualised. In the top panel, the centrality classes are defined with the Glauber MC fit to the UrQMD $N_{ch}$ distribution within the HADES detector acceptance (Glauber-$N_{ch}$) while in the bottom panel, the centrality classes are defined directly on the UrQMD $N_{ch}$ distribution within the HADES detector acceptance (UrQMD-$N_{ch}$). It can be seen that the class boundaries defined in the two cases differ slightly as the Glauber MC fit to UrQMD data is not perfect.

\begin{figure}
    \centering
    \includegraphics[width=0.5\textwidth]{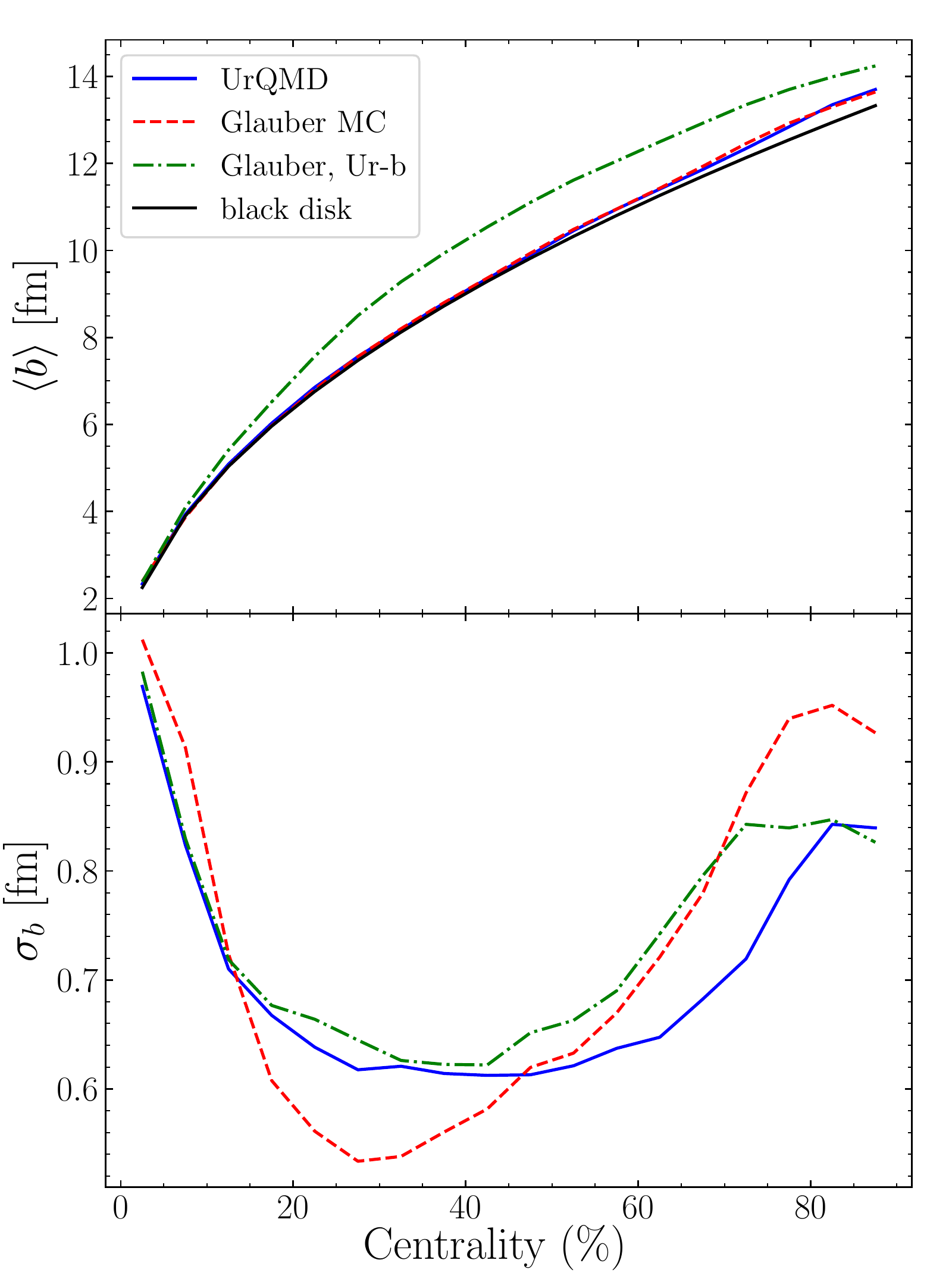}
    \caption{(Color online) Mean and standard deviation of the impact parameter as a function of the centrality class. The top plot shows the mean impact parameter ($\langle b \rangle$), while the bottom plot shows the standard deviation ($\sigma_{b}$) of the impact parameter distributions. The $\langle b \rangle$ and $\sigma_{b}$ from the Glauber MC fit for centrality classes defined on Glauber-$N_{ch}$ distribution is shown as the dashed red curve while the $\langle b \rangle$ and $\sigma_{b}$ defined on the $N_{ch}$ distribution from UrQMD is shown as solid blue curve. The dot-dashed green curve (\emph{Glauber, Ur-b}) depicts the $\langle b \rangle$ and $\sigma_{b}$ taken from UrQMD for $N_{ch}$-centrality-bins defined from the Glauber-$N_{ch}$ distribution. The solid black curve in the top panel shows the impact parameters for different centralities as given by a simplified black disk approximation of the nuclei.}
    \label{bdist}
\end{figure}

\begin{figure}[t]
    \centering
    \includegraphics[width=0.5\textwidth]{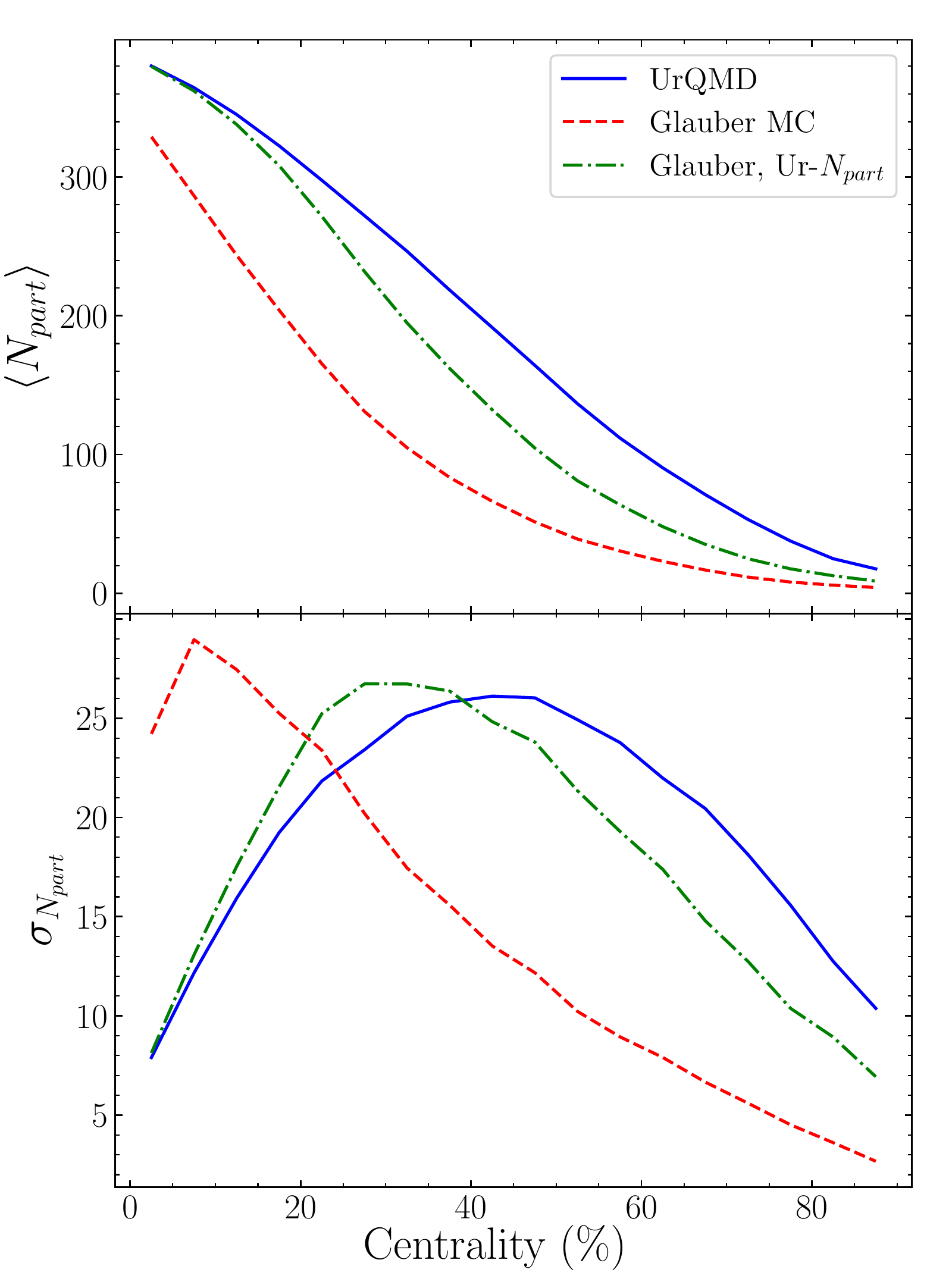}
    \caption{(Color online) Mean and standard deviation of the number of participants as a function of centrality class. The top plot shows the mean ($\langle N_{part} \rangle$) of the distributions of number of participants while the bottom plot shows the standard deviation ($\sigma_{N_{part}}$) of the distributions of number of participants. The $\langle N_{part} \rangle$ and $\sigma_{N_{part}}$ from Glauber MC for centrality classes defined on Glauber-$N_{ch}$ distribution is shown as the dashed red curve while the $\langle N_{part} \rangle$ and $\sigma_{N_{part}}$ from UrQMD for centrality classes defined on UrQMD-$N_{ch}$ distribution is shown as solid blue curve. The dot-dashed green curve (\emph{Glauber, Ur-$N_{part}$}) depicts the $\langle N_{part} \rangle$ and $\sigma_{N_{part}}$ from UrQMD for centrality classes defined on Glauber-$N_{ch}$ distribution. }   
    \label{pdist}
\end{figure}

Using the $N_{ch}$ cuts shown in figure \ref{cenbins}, we can now compare the impact parameter $b$ and $N_{part}$ distributions in the UrQMD and in the Glauber MC for different centrality classes. This is visualised in figures \ref{bdist} and \ref{pdist} where the following three cases are compared.
\begin{enumerate}
    \item $N_{part}$ and $b$ are taken from Glauber MC for centrality classes defined with the Glauber-$N_{ch}$ distribution. These red curves are referred to as \emph{Glauber MC} in figures
    \ref{bdist} and \ref{pdist}.
    \item $N_{part}$ and $b$ are taken from UrQMD and the centrality classes are defined with the true UrQMD-$N_{ch}$ distribution. These blue curves are referred to as \emph{UrQMD} in figures
    \ref{bdist} and \ref{pdist}.
    
    \item $N_{part}$ and $b$ are taken from UrQMD but for $N_{ch}$-bins which have been defined from the Glauber MC fit to the UrQMD distribution. These green curves are referred to as \emph{Glauber, Ur-$b$} and \emph{Glauber, Ur-$N_{part}$}, respectively in figures
    \ref{bdist} and \ref{pdist}. They represent the underlying $b$ and $N_{part}$ distributions in the model which is fit to obtain the Glauber MC centrality classes.
    \end{enumerate}

Figure \ref{bdist} shows that when the impact parameter of an event is consistently taken from the same model as the one used to define the centrality classes, both Glauber MC and UrQMD gives similar impact parameter distributions for each given centrality class. For comparison we also show the impact parameter as function of centrality for a black disc approximation (black solid line). This very simple model also agrees well with both the red and blue lines, except for most peripheral collisions where the assumption of a sharp edge is not valid anymore. As already discussed, the  Glauber MC fit to UrQMD multiplicity distribution had large deviations, especially for intermediate and peripheral collisions. This causes the $N_{ch}$ cuts that define any given centrality class in Glauber-$N_{ch}$ distribution to be different from the cuts defined directly on the UrQMD-$N_{ch}$ distribution. As a result, the average impact parameter extracted from UrQMD using centrality classes as defined with the fitted Glauber-$N_{ch}$ distribution,  differ from the true impact parameter values from UrQMD. The standard deviation of the impact parameter distributions look similar in all three cases for all centralities. The differences in $\sigma_{b}$, for a centrality class, is only about 0.1 fm between the three scenarios. 

\begin{figure}
    \centering
    \includegraphics[width=0.5\textwidth]{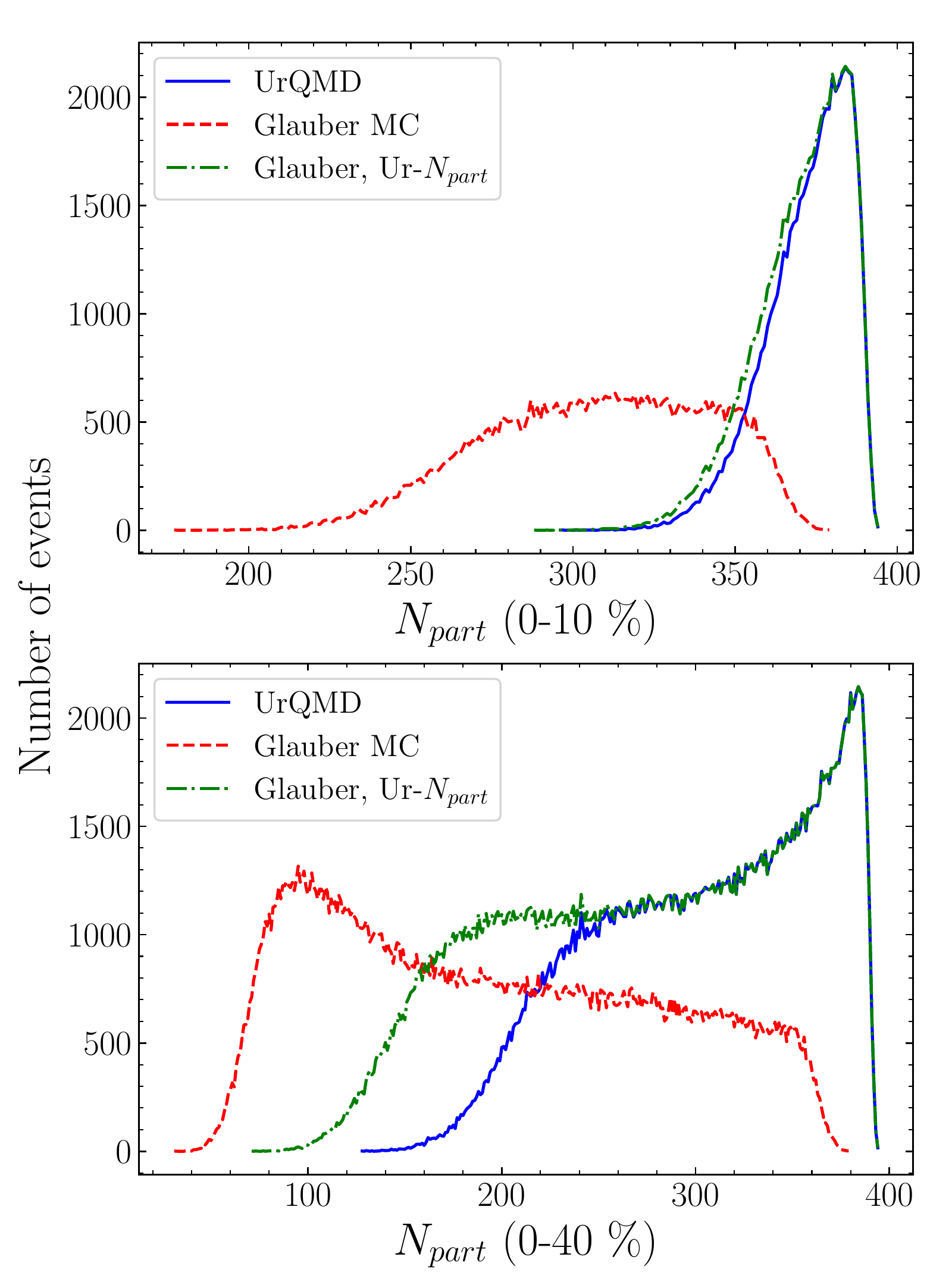}
    \caption{(Color online) $N_{part}$ distributions for 0-10\% (top plot) and 0-40 \% (bottom plot) centrality classes. The $N_{part}$ distribution from Glauber MC for centrality classes defined on Glauber-$N_{ch}$ distribution is shown as the dashed red curve while the $N_{part}$ distribution from UrQMD for centrality classes defined on UrQMD-$N_{ch}$ distribution is shown as solid blue curve. The dot-dashed green curve (\emph{Glauber, Ur-$N_{part}$}) depicts the $N_{part}$ from UrQMD for centrality classes defined on Glauber-$N_{ch}$ distribution. }   
    \label{pdist1}
\end{figure}

However, the results differ drastically for the three scenarios when it comes to the $N_{part}$ distributions. This is demonstrated in figure \ref{pdist}, where the top panel shows the $\langle N_{part} \rangle$ and the bottom plot shows $\sigma_{N_{part}}$ as a function of centrality. For all centrality classes, a huge difference is observed in the number of participants between the Glauber MC and UrQMD models, even when centrality classes are consistently defined using the corresponding model data. The $\langle N_{part} \rangle$ from the Glauber MC for centrality classes defined on the Glauber-$N_{ch}$ distribution are smaller up to a factor of $\approx$2, than the $\langle N_{part} \rangle$ from UrQMD for centrality classes defined on the UrQMD-$N_{ch}$ distribution. The difference between the red and blue curve, in Figure \ref{pdist}, quantifies the error that mostly arises from fitting the Glauber model to a distribution which is based on a different physics model. A significant difference is also observed between the Glauber MC and UrQMD $N_{part}$ distributions, if centrality classes defined using the Glauber-$N_{ch}$ are used. The centrality dependence of the variance of $N_{part}$ distributions also differs drastically between these three cases. The Glauber MC $N_{part}$ distributions have the largest variance ($\sigma_{N_{part}}$ about 30 ) for most central collisions (5-10 \%) when centrality classes are defined from the Glauber-$N_{ch}$ distribution. However, the $N_{part}$ distributions from UrQMD have the largest variance for mid-central collisions (about 40-45 \%) when centrality classes are defined from the UrQMD-$N_{ch}$ distribution.

The differences in the $N_{part}$ distributions for the two models suggests that $N_{part}$ (and its fluctuations) are strongly model dependent, at least at beam energies $\sqrt{s_{\mathrm{NN}}} \lessapprox$ 5-10 GeV. Similar $N_{ch}$ values appear even if the underlying $N_{part}$ distributions are completely different. Figure \ref{pdist1} shows the differences of the $N_{part}$ distributions of Glauber MC and UrQMD for 0-10 \% (top plot) and 0-40 \% (bottom plot) centrality classes. For the 0-10 \% centrality class, the $N_{part}$ distribution from UrQMD for centrality classes defined on the UrQMD-$N_{ch}$ distribution peaks at around 380, while the $N_{part}$ distribution from Glauber MC for centrality classes defined on Glauber-$N_{ch}$ distribution does not show a clear maximum at all. The Glauber-MC $N_{part}$ distribution has huge width and contains events with very small $N_{part}$. For the 0-40 \% centrality class, the UrQMD $N_{part}$ distribution peaks at around 385 while the Glauber $N_{part}$ distribution peaks at around 100, about a factor of 4 lower that the UrQMD value! When centrality classes defined on the Glauber-$N_{ch}$ distribution are applied to UrQMD data, to select the events, the resulting $N_{part}$ distributions are close to the UrQMD $N_{part}$ distributions for centrality classes defined on UrQMD-$N_{ch}$ distribution. This important finding has severe consequences on the volume fluctuations and on its correlations and are relevant for the extraction of proton fluctuations for the HADES experiment and also in the STAR-BES energy range. This is a direct consequence of secondary particle scattering and the contribution of elastic scatterings to the finally observed protons.

 \section{Pion production at 1.23 $A$GeV }
Recently, HADES has published the centrality dependence of charged pion production at a beam energy of 1.23 $A$GeV and the comparison to transport model simulations  \cite{HADES:2020ver} . The pion multiplicity measured by the experiment was found to be lower than the multiplicity predicted by different theoretical models such as PHSD, GiBUU, SMASH, IQMD and PHQMD in given centrality selections. Until now, few ideas have been put forward to explain this discrepancy besides introducing explicitly density dependent production cross sections \cite{Godbey:2021tbt}. 

  \begin{figure}[t]
    \centering
    \includegraphics[width=0.5\textwidth]{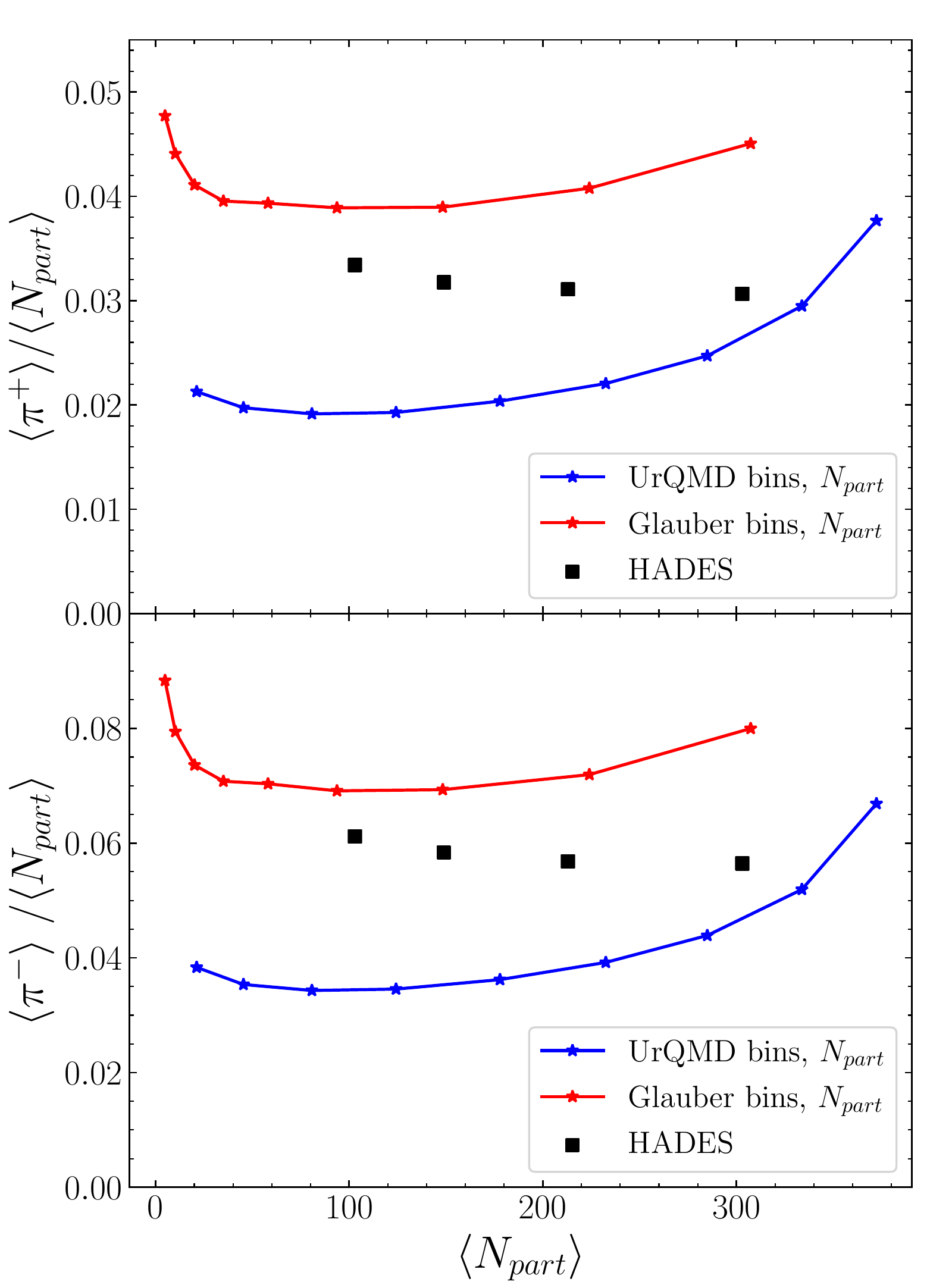}
    \caption{(Color online) Charged pion multiplicity per participant as a function of number of participants. The blue curve (UrQMD bins, $N_{part}$) shows the results for centrality classes defined on UrQMD data with $N_{part}$ also taken from UrQMD. The pion mulitplicity from UrQMD for centrality bins and the number of participants defined by Glauber MC are shown in red (Glauber bins, $N_{part}$). The black squares are the HADES results taken from \cite{HADES:2020ver}.  }   
    \label{7}
\end{figure}

In this section, we report the UrQMD predictions for the charged pion production in 1.23 $A$GeV Au-Au collisions as function of $N_{part}$ and centrality, for different methods of determining them. The charged pion multiplicity per participant predicted by UrQMD is plotted as a function of $\langle N_{part} \rangle$ for different centrality selection methods in figure \ref{7}. As already discussed, different models predict different $N_{part}$ distributions for any given centrality class. Therefore, depending on our choice of model for centrality selection and $\langle N_{part} \rangle$ estimation, the results can vary drastically.  The values for $\langle N_{part} \rangle$ and $\langle \pi \rangle$ used in figure \ref{7} are from 10\% centrality bins for the different models. The last 4 points in all the curves and HADES data in figure \ref{7} therefore correspond to the results from centrality bins 30-40\%, 20-30\%, 10-20\%, 0-10\% respectively. It can be seen that when the Glauber MC model is used for both centrality selection and to estimate the $\langle N_{part} \rangle$, the resulting pion multiplicity per participant from UrQMD is larger than the HADES measurements at all centralities (red curve). However, when the UrQMD model is used for both centrality selection and to estimate the $\langle N_{part} \rangle$, the pion multiplicity per participant from UrQMD found to be smaller than the HADES measurements for mid-central and peripheral collisions (blue curve) and larger than the HADES data for the most central collisions (0-10 \%). Due to the large differences in $\langle N_{part} \rangle$ for each of these centrality classes for the two models, it is not advised to study the pion multiplicity as a function of $\langle N_{part} \rangle$. The $\langle N_{part} \rangle$ is not a reliable quantity to compare experimental observations such as particle yields with models due to the large differences between the Glauber MC assumption made by the experiments and the distributions simulated in transport models. 
  
\begin{figure}
    \centering
    \includegraphics[width=0.5\textwidth]{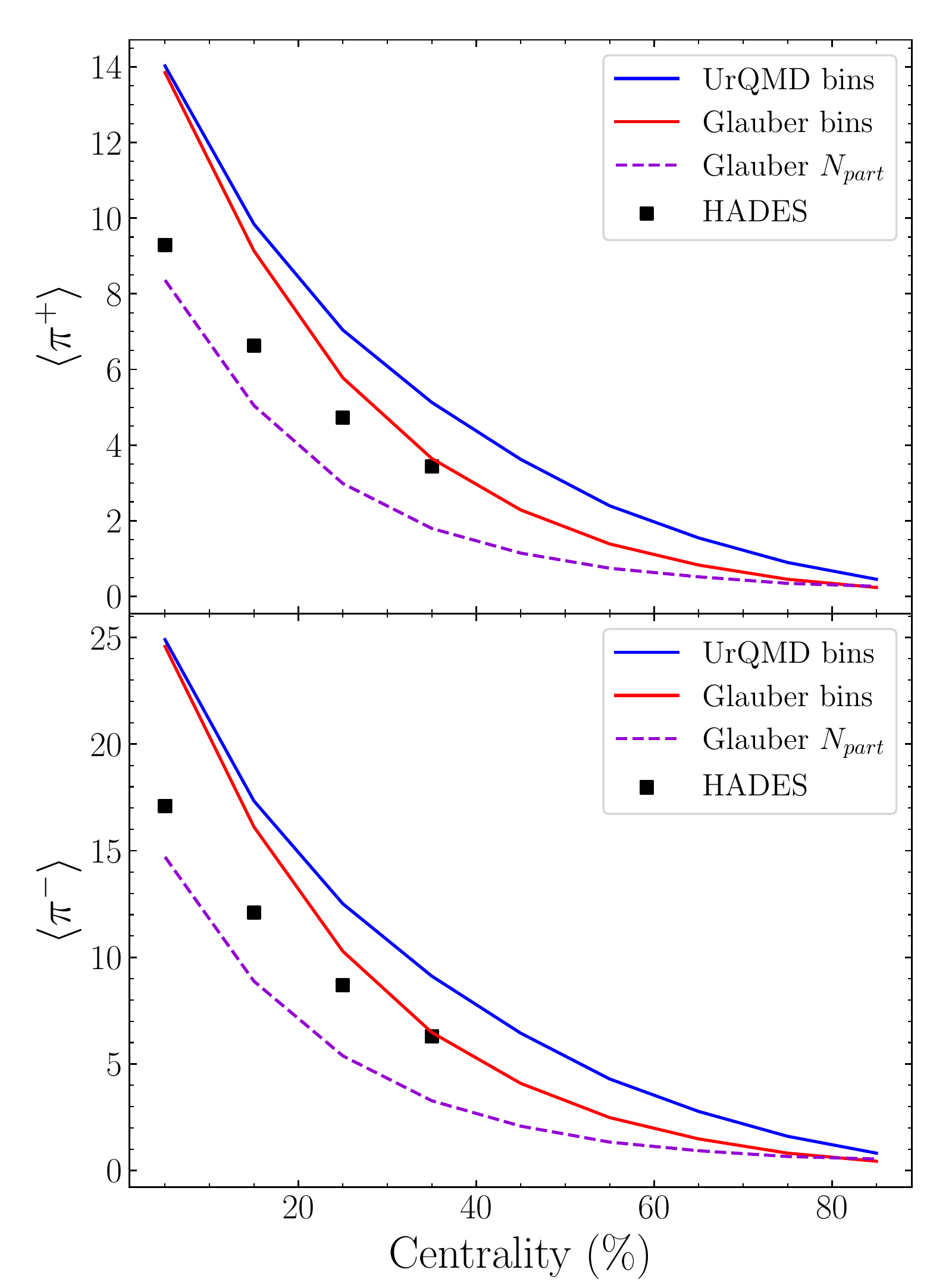}
    \caption{(Color online) Charged pion multiplicities as a function of centrality. The plots depicts the $\langle \pi^+ \rangle$ and $\langle \pi^- \rangle$ multiplicities from UrQMD and HADES. The curves "Glauber bins" and "UrQMD bins" are the results from UrQMD for Glauber-$N_{ch}$ and UrQMD-$N_{ch}$ based centrality selections respectively.The HADES results (black squares) are taken from \cite{HADES:2020ver}. The dashed violet curve shows the pion multiplicities from UrQMD when the events are sampled to have $N_{part}$ distribution similar to the Glauber MC predictions for any given centrality class. }   
    \label{6}
\end{figure}
 
The ambiguities associated with $\langle N_{part} \rangle$ can be avoided if we analyse the multiplicities as a function of centrality, defined consistently by a percentile of $N_{ch}$ from either a Glauber MC fit or directly the model simulation. The charged pion multiplicities from UrQMD is plotted as a function of centrality in figure \ref{6}. In the figure, the results from UrQMD for centrality classes defined from the UrQMD-$N_{ch}$ distribution is plotted in blue ("UrQMD bins") while the UrQMD results for Glauber-$N_{ch}$ based centrality selection is plotted in red ("Glauber bins"). The "Glauber bins" curve should match the "UrQMD bins" curve if the Glauber MC fit to the UrQMD charged particle multiplicity distribution was perfect. It can be seen that UrQMD predicts significantly larger pion multiplicities for all centrality classes than what is measured by HADES in this case. The results are similar to the predictions from other theoretical models, as shown in \cite{HADES:2020ver}.

 It is important to note that, when comparing experimental observations with theoretical models, it is usually assumed that for events with similar centrality, the underlying $N_{part}$ distributions are also similar. However, as we have shown, this is not at all the case. Even though the Glauber MC model can be fit to an experimental or theoretical $N_{ch}$ distribution, the $N_{part}$ distribution as extracted from a Glauber MC for a centrality class can be completely different from the true $N_{part}$ distribution. To demonstrate the uncertainty this introduces in the analysis, instead of using the centrality cuts in $N_{ch}$ to select events, they are sampled to have the $N_{part}$ distribution similar to that predicted by Glauber MC for any given centrality class. The results are shown as the dashed violet curves in figure \ref{6}. It can be seen that unlike the UrQMD-$N_{ch}$ based event selection which predicted more pions than the HADES observations, when the sampled events have Glauber $N_{part}$ distribution for each centrality class, the predicted pion multiplicities are lower than the HADES data. The strong centrality dependence in the deviation from HADES data is also reduced in this case. This shows that when comparing the data to a theoretical model, there will be a large uncertainty in the predicted pion multiplicity if the assumption of the underlying $N_{part}$ distribution is incorrect or at least not fully consistent.

 \section{Further discussions on the definition of $N_{part}$}

 So far, the discussion of $N_{part}$ in the UrQMD model has followed a straightforward definition, where it includes all nucleons from the incoming nuclei that experience at least one collision. While this is a reasonable assumption due to the usually significant momentum transfer, it may still be interesting to discuss a slightly different definition of $N_{part}$ from the UrQMD model which may be able to mimic a Glauber model. In such a scenario, which we will refer to as "\textit{UrQMD$-$inel}" in the following, we only count nucleons which have undergone at least one inelastic scattering as participants. This means that we consider nucleons with any arbitrary number of scatterings as spectators, as long as these scatterings were elastic. This approach takes into account secondary interactions but also nucleons that end up in the experiments' acceptance as participants may be counted as spectators. To be clear, we do not claim that such a definition of participants is physically more meaningful than the usual definition, however it may help in understanding possible systematic uncertainties that stem from such different definitions when models are compared to data.

 \begin{figure}[t]
    \centering
    \includegraphics[width=0.5\textwidth]{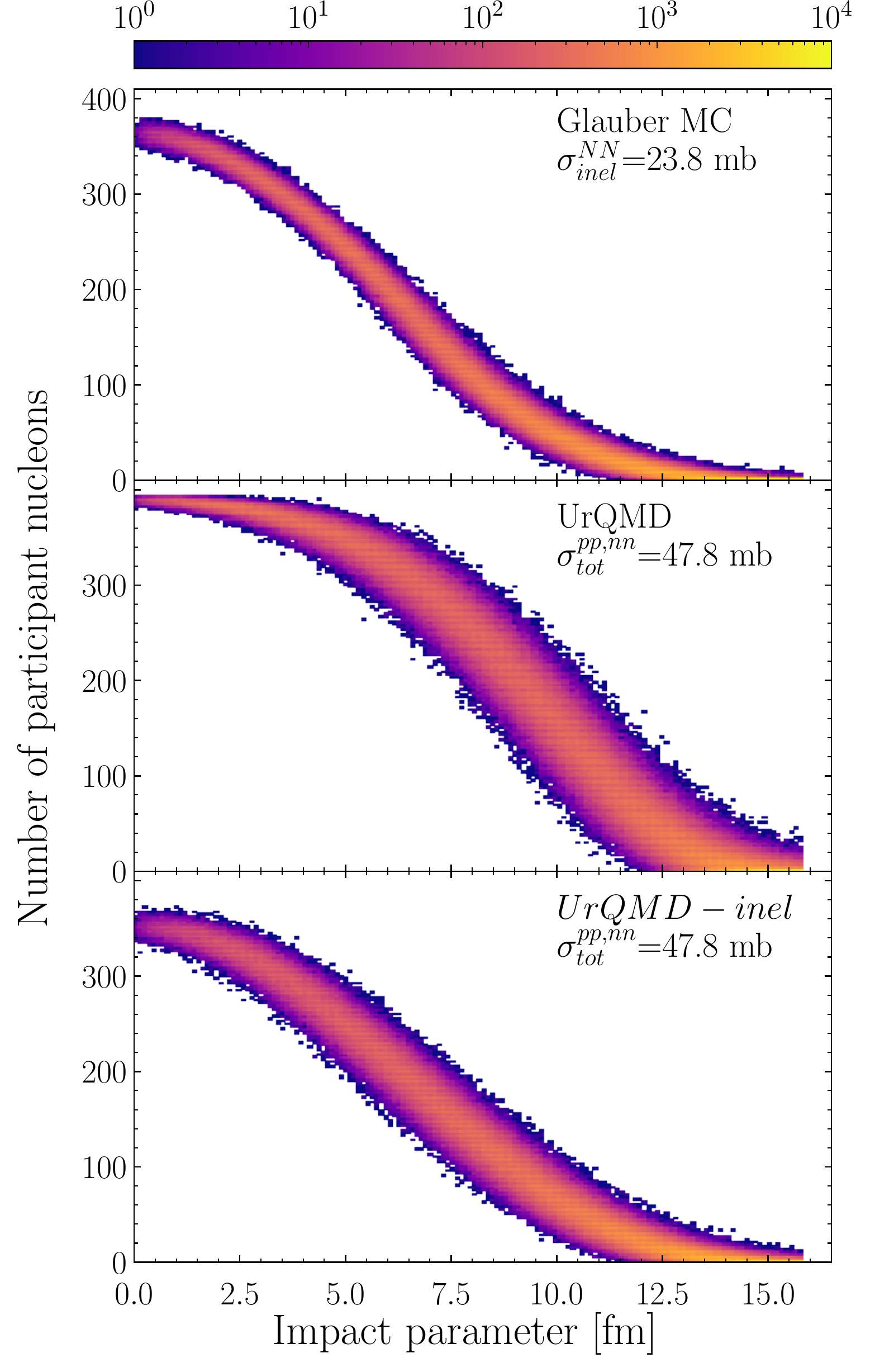}
    \caption{(Color online) Distributions of the number of participants as function of the impact parameter for the Glauber MC model (upper panel), the default UrQMD model with the default definition of a participant (middle panel) and the new definition \textit{UrQMD$-$inel} (lower panel). The first two results are the same as shown in the top and bottom panels in figure \ref{4}. The new \textit{UrQMD$-$inel} results are much more similar to the Glauber MC, yet show a broader spread in the $N_{part}$ for a given impact parameter. }  
    \label{np-bnew}
\end{figure}

Figure \ref{np-bnew} shows the distribution of the number of participants as function of the impact paramater from the Glauber MC model (upper panel), compared to the usual definition (UrQMD, in the middle panel) and the alternate definition (\textit{UrQMD$-$inel} in the lower panel). What can be clearly observed is that when only nucleons with at least one inelastic scattering are counted as participants, the number of participants decreases and for a given impact parameter, the average number becomes similar to that of the Glauber model. This is not surprising as by this definition, a lot of secondary elastic interactions are disregarded similar to the Glauber assumption. However, for a given impact parameter, the distribution of $N_{part}$ remains broader than in the Glauber model.

\begin{figure}[t]
    \centering
    \includegraphics[width=0.5\textwidth]{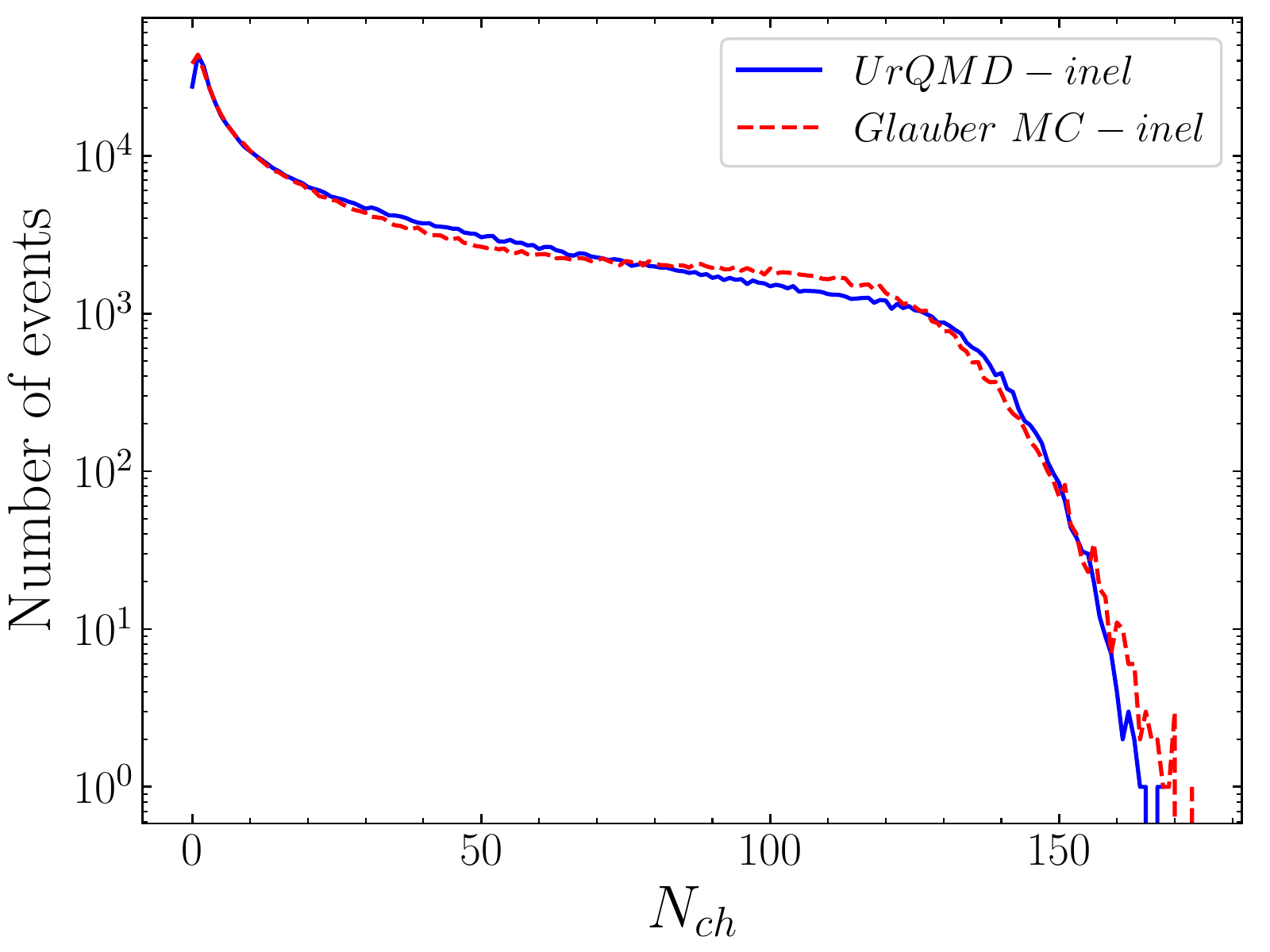}
    \caption{(Color online) UrQMD distribution of the charged particle multiplicity for $6\times 10^{5}$ inelastic events based on the "\textit{UrQMD$-$inel}" definition of a participant (blue solid line) together with a Glauber MC fit to the model results (red dashed line). The fit quality is much improved as compared to the Glauber fit to the default UrQMD curve shown in figure \ref{0}.}   
    \label{fitnew}
\end{figure}

As the distribution of the number of participants is now closer to the Glauber model, fitting the $N_{ch}$ distribution of the UrQMD model, again neglecting charged nucleons without any inelastic collisions, with the Glauber MC approach, yields a much better fit. This new fit is now shown in figure \ref{fitnew}.

\begin{figure}[t]
    \centering
    \includegraphics[width=0.5\textwidth]{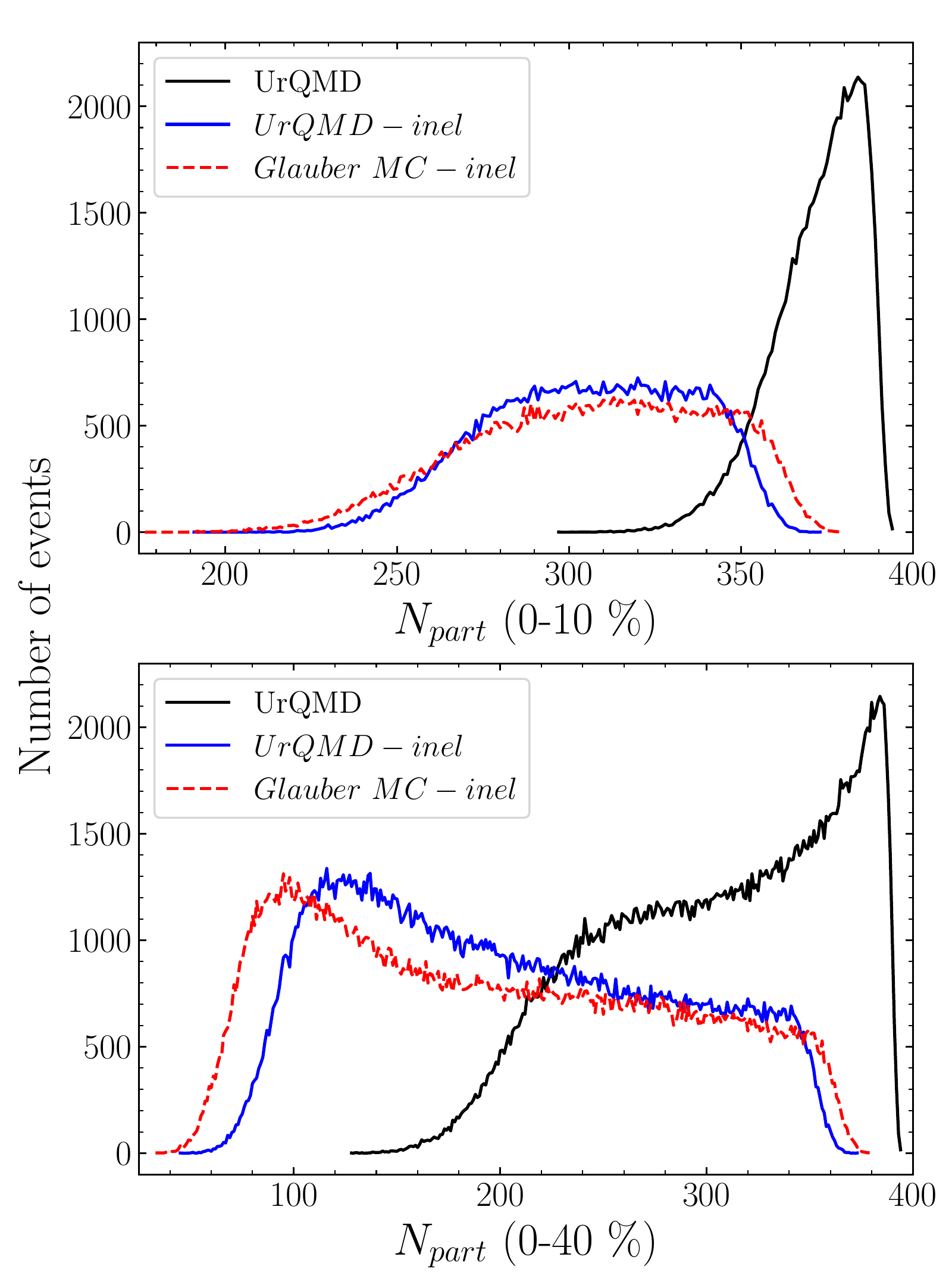}
    \caption{(Color online) Distributions of the number of participants $N_{part}$ for 0-10 $\%$ (upper figure) and 0-40 $\%$ most central (lower panel) collisions. The centrality bins are either defined by percentiles of the $N_{ch}$ distributions directly from the UrQMD model (solid lines) or from a Glauber MC fit to the model results (dashed line). The solid blue line, corresponding to the new \textit{"inel"} definition of a participant is very different from the default definition. It is also much closer to the Glauber MC fit which is in stark contrast to figure \ref{pdist1}.}
    \label{flucnew}
\end{figure}

A strong effect of the new definition of a participant can be observed if one looks at the fluctuations of $N_{part}$. These are shown in figure \ref{flucnew} for centrality classes 0-10\% (upper panel) and 0-40\% (lower panel). The distributions were obtained, as before, by defining the centrality classes to be percentiles of the charged particle multiplicity distribution in the simulations. As one can see, the \textit{UrQMD$-$inel} results resemble very closely the Glauber MC results and are very different to the UrQMD results when all scattered nucleons are counted.

These results show that it is possible to easily mimic the Glauber MC model within UrQMD which makes it possible to understand the systematic uncertainties of these two different centrality definitions when comparing to experimental data that uses the Glauber model for centrality definition. It should be stated however, that we do not claim that the \textit{UrQMD$-$inel} results are in any way physically motivated. One may argue that few of the elastically scattered nucleons may be reabsorbed in the spectator fragments but many of them end up in phase space regions where they can and will be measured by the experiment. This highlights again how problematic it is to compare model simulations with data based on centrality definitions that rely on certain model assumptions, be it the Glauber model or the UrQMD model.

\section{Consistency test for $N_{part}$}

 \begin{figure}[t]
    \centering
    \includegraphics[width=0.5\textwidth]{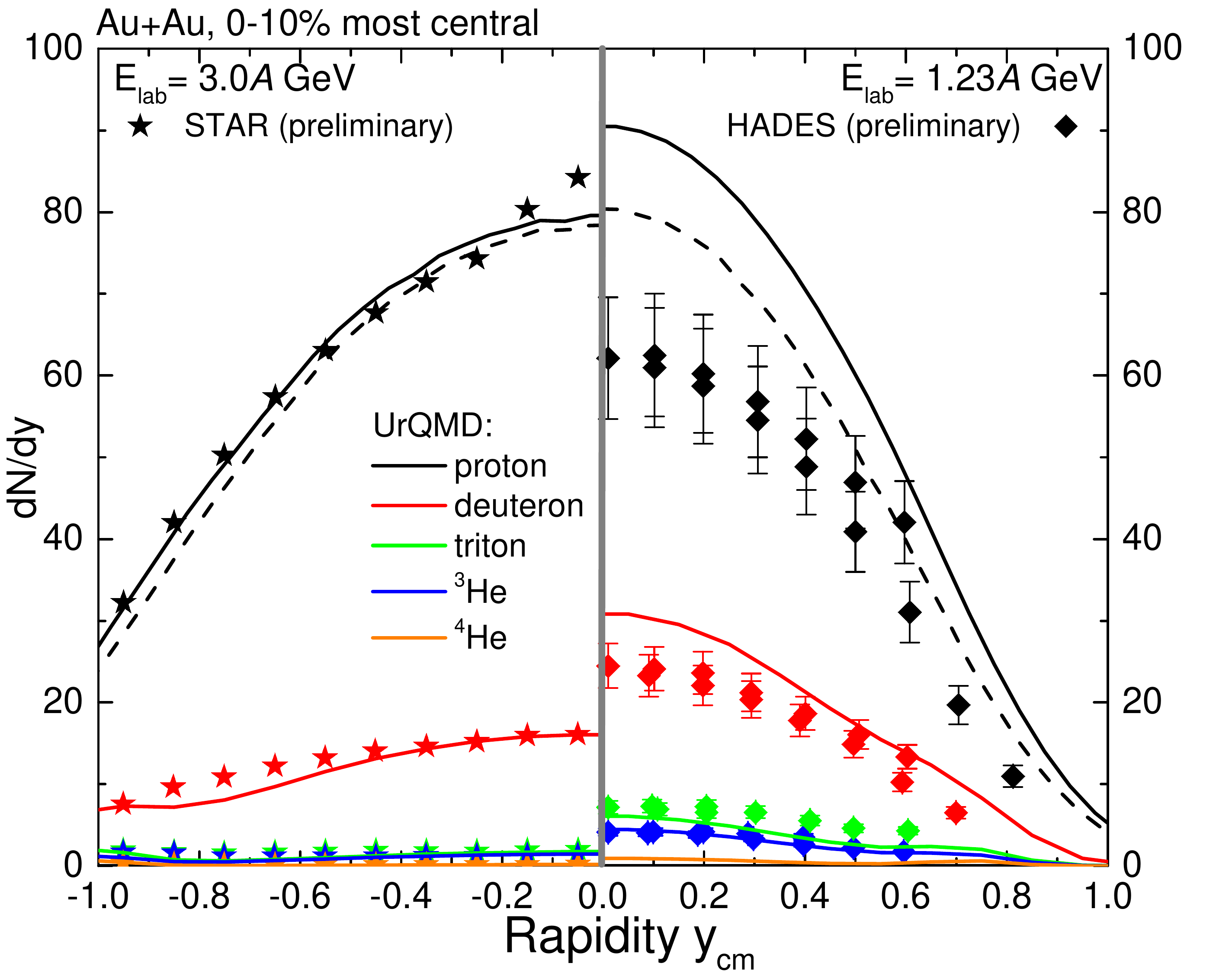}
    \caption{(Color online) 
     Rapidity distributions of free protons and light nuclei in the most central collisions of Au-Au nuclei at two different beam energies from the UrQMD model plus coalescence (solid lines). The dashed lines correspond to all protons that have at least scattered inelastically once, consistent with the \textit{UrQMD$-$inel} description. Shown are either the forward or backward hemisphere only, due to symmetry reasons. The simulations are compared to the preliminary data from the two respective experiments STAR \cite{Liu:2022ump} (left, stars) and HADES \cite{Schuldes:2016eqz,HADES:lorenzd} (right, diamonds) shown as symbols. While the STAR data are well reproduced, the HADES data appear to show significantly fewer protons and light nuclei at all rapidities (note that the beam rapidity in this frame for STAR is $y_{beam}\approx 1.05$ and for HADES is  $y_{beam}\approx 0.75$). The sum of all participating baryons minus the free neutrons is $\approx 220$ for both cases within UrQMD, while for HADES it is $\approx 190$.}
    \label{dndy}
\end{figure}

Up to now, the role of $N_{part}$ has been discussed as a model dependent, and likely unreliable, quantity. However, there may be actual observable consequences related to $N_{part}$ in the beam energy range under discussion. As the total baryon number in the system is conserved and rarely any baryon-antibaryon pairs are produced, the number of participating baryons could be measured directly. In fact the only observation missing is that of the \emph{free} neutrons, $N_n$, which cannot be detected by the current experiments which measure only charged particles. Besides that, data on the rapidity distribution of the free protons and baryons in light nuclei is available (albeit preliminary) from the HADES and STAR experiments.  These can be contrasted with our UrQMD simulations which has been extended to include a coalescence mechanism to describe light nuclei production \cite{Hillmann:2021zgj}. The comparison between the UrQMD simulations for the $0-10 \%$ most central (defined by impact parameter $b<4.7$ fm) is shown in figure \ref{dndy}. As one can see the light nuclei give a relevant contribution to the total baryon number. What is also visible is that the UrQMD model provides a very good description of all STAR rapidity distributions (figure \ref{dndy} left) except at rapidities close to the fragmentation region. At the same time the sum of all measured baryons in this case is $N_{part}- N_n \approx 220 $. The same result on the UrQMD observable baryon number is also obtained at the HADES energy (figure \ref{dndy} right). Adding the number of neutrons both results are consistent with the total $N_{part}$ discussed earlier. The HADES data on the other hand show a clear suppression of proton and deuteron rapidity spectra compared to the model simulation. In turn this leads to a HADES  $N_{part}- N_n \approx 190$ which is surprisingly in agreement with the Glauber model number of participants. This time however, this discrepancy also results in an observable discrepancy mostly of the free proton $dN/dy$ at mid-rapidity which cannot be explained by a simple redefinition. 

The black dashed lines indicate the proton distributions when only such protons are selected which have undergone at least one inelastic scattering. It is found that this method does not allow to fully explain the difference in actually measured proton number in HADES. For the STAR results at 3 \textit{A}GeV, almost no difference between the solid and dashed black lines is observed. This indicates that especially in the beam energy range of 1-2 $A$GeV the description of the initial compression phase is highly non-trivial.

If this discrepancy between experiments and between simulation and experiment remains, this needs further investigation. There are several possible apparent reasons for this observation. For example, this effect can stem from the different acceptance of the HADES and STAR experiments which may prevent the detection of some of the protons, or there is a systematic incompatibility in the centrality determination of either experiment. A sharp change in the underlying physics, modelled by the transport description, is also possible and would likewise warrant more detailed studies.

\section{Conclusion}

The Glauber MC and UrQMD models are investigated in detail and the differences in their predictions are presented.  It is shown that the Glauber MC predicts completely different $N_{part}$ distribution in comparison to the microscopic model UrQMD. The fact that the Glauber MC can be fit to the "experimental data" does not assure that the data shares the same underlying $N_{part}$ distribution as that predicted by the Glauber MC model. This strong model dependence of $N_{part}$ can skew the results of experimental analysis if $N_{part}$ used is not consistent throughout the analysis and the final model-to-data comparison. Our study showed what happens if the Glauber model is applied as intended on a model with different physics assumptions. To demonstrate this effect, we studied the charged pion production and the proton and light nuclei rapidity spectra at $E_{\mathrm{lab}}$ = 1.23 \textit{A}GeV and compared the results to HADES measurements. We showed that the pion multiplicity per participant can be highly sensitive to the model used to estimate the underlying $N_{part}$ distribution for a given centrality class. On the other hand,
it is shown that the impact parameter and centrality classes, extracted from a Glauber fit, when done consistently with UrQMD output, actually very well agrees with the true UrQMD results. It is suggested that studying observables such as the pion multiplicity as a function of centrality instead of $N_{part}$ can partly avoid the explicit model dependency of $N_{part}$ that arises when comparing experimental data to model predictions. Nevertheless, this also does not assure that the experimental data and the model being compared share similar $N_{part}$ distribution.

It was also shown that the model dependence of $N_{part}$ has measurable consequences in the observed rapidity distributions of free protons and light nuclei. Here, the preliminary data of the HADES experiment also show a significant reduction of the proton $dN/dy$ as compared to UrQMD, consistent with a significantly smaller $N_{part}$, an effect which was not observed e.g. in data from the STAR experiment. Apparently, understanding these inconsistencies is important also in the interpretation of observables like the pion multiplicity and proton number fluctuations. One possible solution to this issue is to explicitly measure the $N_{part}$ in experiments using detectors such as a Zero Degree Calorimeter and then compare the model results for events with the same $N_{part}$ distribution. 

It would also be interesting to investigate if one can accurately extract $N_{part}$ based on models like UrQMD using machine learning or deep learning methods. It has already been shown in \cite{OmanaKuttan:2020brq,OmanaKuttan:2021axp,OmanaKuttan:2020btb} that one can accurately and consistently map the impact parameter of the collision, or the underlying Equation-of-State that is defined in a model, to the experimental output. It would be worthwhile to investigate the possibility to extend such methods to extract a consistently defined $N_{part}$ from experimental data. However, such investigations are beyond the scope of this paper and are desirable for future research. 

\begin{acknowledgement}
The authors thank B.~Kardan for help with the Glauber MC model as well as M.~Lorenz and J.~Stroth for helpful discussions about the HADES data. M.O.K. thanks the GSI and HFHF as well as the SAMSON AG for their support. K.Z. and J.S. thank the Samson AG and the BMBF through the ErUMData project for funding. H.S. acknowledges the Walter Greiner Gesellschaft zur F\"orderung der physikalischen Grundlagenforschung e.V. through the Judah M. Eisenberg Laureatus Chair at Goethe Universit\"at Frankfurt am Main. The computational resources for this project where provided by the Frankfurt Center for Scientific computing through the Goethe-HLR cluster.  

\end{acknowledgement}
\bibliographystyle{spphys}
\bibliography{main.bib}

\begin{thebibliography}{10}
\providecommand{\url}[1]{{#1}}
\providecommand{\urlprefix}{URL }
\expandafter\ifx\csname urlstyle\endcsname\relax
  \providecommand{\doi}[1]{DOI \discretionary{}{}{}#1}\else
  \providecommand{\doi}{DOI \discretionary{}{}{}\begingroup
  \urlstyle{rm}\Url}\fi

\bibitem{Stoecker:1986ci}
H.~Stoecker, W.~Greiner, Phys. Rept. \textbf{137}, 277 (1986).
\newblock \doi{10.1016/0370-1573(86)90131-6}

\bibitem{Hofmann:1976dy}
J.~Hofmann, H.~Stoecker, U.W. Heinz, W.~Scheid, W.~Greiner, Phys. Rev. Lett.
  \textbf{36}, 88 (1976).
\newblock \doi{10.1103/PhysRevLett.36.88}

\bibitem{Stoecker:2004qu}
H.~Stoecker, Nucl. Phys. A \textbf{750}, 121 (2005).
\newblock \doi{10.1016/j.nuclphysa.2004.12.074}

\bibitem{Stephanov:1998dy}
M.A. Stephanov, K.~Rajagopal, E.V. Shuryak, Phys. Rev. Lett. \textbf{81}, 4816
  (1998).
\newblock \doi{10.1103/PhysRevLett.81.4816}

\bibitem{Hatta:2003wn}
Y.~Hatta, M.A. Stephanov, Phys. Rev. Lett. \textbf{91}, 102003 (2003).
\newblock \doi{10.1103/PhysRevLett.91.102003}.
\newblock [Erratum: Phys.Rev.Lett. 91, 129901 (2003)]

\bibitem{Glauber:1955qq}
R.J. Glauber, Phys. Rev. \textbf{100}, 242 (1955).
\newblock \doi{10.1103/PhysRev.100.242}

\bibitem{Glauber:2006gd}
R.J. Glauber, Nucl. Phys. A \textbf{774}, 3 (2006).
\newblock \doi{10.1016/j.nuclphysa.2006.06.009}

\bibitem{Baumgardt:1975qv}
H.G. Baumgardt, J.U. Schott, Y.~Sakamoto, E.~Schopper, H.~Stoecker, J.~Hofmann,
  W.~Scheid, W.~Greiner, Z. Phys. A \textbf{273}, 359 (1975).
\newblock \doi{10.1007/BF01435578}

\bibitem{Cimerman:2023hjw}
J.~Cimerman, I.~Karpenko, B.~Tomasik, P.~Huovinen, arXiv:2301.11894  (2023)

\bibitem{Reichert:2023eev}
T.~Reichert, O.~Savchuk, A.~Kittiratpattana, P.~Li, J.~Steinheimer,
  M.~Gorenstein, M.~Bleicher, arXiv:2302.13919  (2023)

\bibitem{Stoecker:1980vf}
H.~Stoecker, J.A. Maruhn, W.~Greiner, Phys. Rev. Lett. \textbf{44}, 725 (1980).
\newblock \doi{10.1103/PhysRevLett.44.725}

\bibitem{Brachmann:1999xt}
J.~Brachmann, S.~Soff, A.~Dumitru, H.~Stoecker, J.A. Maruhn, W.~Greiner, L.V.
  Bravina, D.H. Rischke, Phys. Rev. C \textbf{61}, 024909 (2000).
\newblock \doi{10.1103/PhysRevC.61.024909}

\bibitem{Ohnishi:2017xjg}
A.~Ohnishi, Y.~Nara, H.~Niemi, H.~Stoecker, Acta Phys. Polon. Supp.
  \textbf{10}, 699 (2017).
\newblock \doi{10.5506/APhysPolBSupp.10.699}

\bibitem{HADES:2022osk}
J.~Adamczewski-Musch, et~al., arXiv:2208.02740  (2022)

\bibitem{Bass:1998ca}
S.A. Bass, et~al., Prog. Part. Nucl. Phys. \textbf{41}, 255 (1998).
\newblock \doi{10.1016/S0146-6410(98)00058-1}

\bibitem{Bleicher:1999xi}
M.~Bleicher, et~al., J. Phys. G \textbf{25}, 1859 (1999).
\newblock \doi{10.1088/0954-3899/25/9/308}

\bibitem{Petersen:2008dd}
H.~Petersen, J.~Steinheimer, G.~Burau, M.~Bleicher, H.~St\"ocker, Phys. Rev. C
  \textbf{78}, 044901 (2008).
\newblock \doi{10.1103/PhysRevC.78.044901}

\bibitem{HADES:2009aat}
G.~Agakishiev, et~al., Eur. Phys. J. A \textbf{41}, 243 (2009).
\newblock \doi{10.1140/epja/i2009-10807-5}

\bibitem{HADES:2017def}
J.~Adamczewski-Musch, et~al., Eur. Phys. J. A \textbf{54}(5), 85 (2018).
\newblock \doi{10.1140/epja/i2018-12513-7}

\bibitem{HADES:2020wpc}
J.~Adamczewski-Musch, et~al., Phys. Rev. C \textbf{102}(2), 024914 (2020).
\newblock \doi{10.1103/PhysRevC.102.024914}

\bibitem{STAR:2022qmt}
STAR, arXiv:2209.11940  (2022)

\bibitem{behruzthesis}
B.~Kardan, {Centrality Determination at 1.23 AGeV Gold-Gold collision and
  readout-electronics for the HADES electromagnetic calorimeter}.
\newblock Master's thesis, Johann Wolfgang Goethe-Universität, Frankfurt am
  Main, Germany (2015)

\bibitem{Workman:2022ynf}
R.L. Workman, et~al., PTEP \textbf{2022}, 083C01 (2022).
\newblock \doi{10.1093/ptep/ptac097}

\bibitem{HADES:2020ver}
J.~Adamczewski-Musch, et~al., Eur. Phys. J. A \textbf{56}(10), 259 (2020).
\newblock \doi{10.1140/epja/s10050-020-00237-2}

\bibitem{Godbey:2021tbt}
K.~Godbey, Z.~Zhang, J.W. Holt, C.M. Ko, Phys. Lett. B \textbf{829}, 137134
  (2022).
\newblock \doi{10.1016/j.physletb.2022.137134}

\bibitem{Liu:2022ump}
H.~Liu, in \emph{{29th International Conference on Ultra-relativistic
  Nucleus-Nucleus Collisions}} (2022)

\bibitem{Schuldes:2016eqz}
H.~Schuldes, {Charged kaon and $\phi$ reconstruction in Au+Au collisions at
  1.23 AGeV}.
\newblock Ph.D. thesis, Goethe U., Frankfurt (main) (2016)

\bibitem{HADES:lorenzd}
M.~Lorenz,   (at SQM 2019)

\bibitem{Hillmann:2021zgj}
P.~Hillmann, K.~K\"afer, J.~Steinheimer, V.~Vovchenko, M.~Bleicher, J. Phys. G
  \textbf{49}(5), 055107 (2022).
\newblock \doi{10.1088/1361-6471/ac5dfc}

\bibitem{OmanaKuttan:2020brq}
M.~Omana~Kuttan, J.~Steinheimer, K.~Zhou, A.~Redelbach, H.~Stoecker, Phys.
  Lett. B \textbf{811}, 135872 (2020).
\newblock \doi{10.1016/j.physletb.2020.135872}

\bibitem{OmanaKuttan:2021axp}
M.~Omana~Kuttan, J.~Steinheimer, K.~Zhou, A.~Redelbach, H.~Stoecker, Particles
  \textbf{4}(1), 47 (2021).
\newblock \doi{10.3390/particles4010006}

\bibitem{OmanaKuttan:2020btb}
M.~Omana~Kuttan, K.~Zhou, J.~Steinheimer, A.~Redelbach, H.~Stoecker, JHEP
  \textbf{21}, 184 (2020).
\newblock \doi{10.1007/JHEP10(2021)184}

\end{thebibliography}

\end{document}